\documentclass[amsmath,amssymb,twocolumn]{revtex4-1}

\usepackage{amsmath,blindtext,booktabs,eurosym,graphicx,hyperref, mathtools}
\usepackage[utf8]{inputenc}
\usepackage[varg]{txfonts}

\newcommand{\ii}{\mathrm{i}}
\newcommand{\e}{\mathrm{e}}
\renewcommand{\d}{\mathrm{d}}

\newcommand{\tens}[1]{\mathbf{#1}}
\newcommand{\cvector}[1]{\left(\begin{array}{c}#1\end{array}\right)}
\renewcommand{\matrix}[2]{\left(\begin{array}{#1}#2\end{array}\right)}

\newcommand{\scalar}[2]{\left\langle\mathbf{#1},\mathbf{#2}\right\rangle}

\newcommand{\usi}[1]{\int \mathrm{d} #1 \,}				
\newcommand{\der}[2]{\frac{\delta{#1}}{\ii \delta{#2}}}		

\newcommand{\nexp}[1]{\mathrm{exp}\left\{ #1 \right\}}

\DeclareMathOperator{\dirac}{\delta_\textsc{D}}
\newcommand {\IC}{\int\d\Gamma_{\ii}}

\begin{document}
\title{Non-equilibrium statistical field theory for classical particles: Impact of correlated initial conditions on non-ideal gases}
\author{Elena Kozlikin, Felix Fabis, Robert Lilow, Celia Viermann, Matthias Bartelmann}
\affiliation{Heidelberg University, Zentrum f\"ur Astronomie, Institut f\"ur Theoretische Astrophysik, Philosophenweg 12, 69120 Heidelberg, Germany}

\begin{abstract}
We use the non-equilibrium statistical field theory for classical particles recently developed by \citeauthor{2011PhRvE..83d1125M} and \citeauthor{2012JSP...149..643D}, together with the free generating functional for particles initially correlated in phase space derived in \citeauthor{paper_1} to study the impact of initial correlations on the equation of state of real gases. We first show that we can reproduce the well known van der Waals equation of state for uncorrelated initial conditions using this approach. We then impose correlated initial conditions and study their qualitative and quantitative effect on the equation of state of a van der Waals gas. The correlations impose a significant correction to the pressure of an ideal gas which is an order of magnitude larger than the correction due to particle interactions. 
\end{abstract}

\maketitle
\section{Introduction}
In a recent paper by \citeauthor{paper_1} \cite{paper_1} a free generating functional for canonical ensembles of microscopic classical particles whose positions and momenta are initially correlated in phase space was derived within the framework of a non-equilibrium field theory for classical particles, building upon the pioneering works of \citeauthor{2011PhRvE..83d1125M} and \citeauthor{2013JSP...152..159D} \cite{2013JSP...152..159D, 2012JSP...149..643D, 2011PhRvE..83d1125M, 2010PhRvE..81f1102M}.\\
As mentioned in \cite{paper_1} this approach can be used for a broad variety of systems. In this work we study a general non-ideal gas with short ranged interactions in a finite volume without any further restrictions. We are looking to provide a qualitative and quantitative analysis of the impact of initial phase-space  correlations on non-ideal gases.\\
Considering gases and their behaviour the usual approach is to set up an equation of state which relates the density to the pressure and the temperature. The changes introduced by initial phase-space correlations into the equation of state are not obvious at first glance and require a detailed analysis to provide some intuition for the behaviour of the system under consideration.\\
We begin with a brief summary of the theory and derive an expression for the full generating functional to first order in the particle interactions using the Mayer cluster expansion in Sect. \ref{sec:fullfunc}.  We use this expression in Sect. \ref{sec:vanderWaals} to derive the van der Waals equation of state as a consistency check for the theory. In  Sect. \ref{sec:corrinitialcond} we finally compute the equation of state for an initially correlated non-ideal gas and study its behaviour in comparison with an uncorrelated van der Waals gas. In Sect. \ref{sec:conclusion} we conclude with a brief summary of our results.\\
Our initial motivation for this study was to understand the effect correlated initial conditions have on cosmic structure formation. This approach, however, can be used for a broad variety of systems whenever correlated initial conditions are involved and may prove useful to describe such effects as phase transitions in initially correlated systems. We will therefore keep our approach as general as possible using simple toy models whenever it is appropriate to demonstrate certain effects.
\section{Fully interacting generating functional}\label{sec:fullfunc}
\subsection{Brief summary of the theory}
We start out with the perturbation ansatz with respect to particle trajectories for the free generating functional for a canonical ensemble introduced by \citeauthor{paper_1} \cite{paper_1}
\begin{equation}
	Z_0[H,\tens{J},\tens{K}]=\e^{\ii H\cdot\hat\Phi}Z_0[\tens J,\tens K]
	\label{eq:genfunc}
\end{equation} 
where $Z_0[\tens J,\tens K]$ is the free generating functional
\begin{equation}
	Z_0[\tens J,\tens K]=\IC{ \, \nexp{\ii \usi t \left\langle{\tens J(t),\tens{\bar x}(t)}\right\rangle}}
\end{equation}
and the term $H\cdot\hat\Phi$ is understood to abbreviate
\begin{equation}
  H\cdot\hat\Phi = \sum_{a=\rho, B}\int\d 1\,H_a(1)\,\hat\Phi_a(1)\;.
\end{equation} 
The phase-space coordinates of the complete particle ensemble,
\begin{equation}
  \vec x_j \coloneqq \cvector{\vec q_j \\ \vec p_j}\;,\quad
  \tens x \coloneqq \vec x_j\otimes\vec e_j\;,
\end{equation}
are bundled into a tensorial structure where the Einstein convention is used to sum over repeated indices and where $\vec e_j$ is the $N$-dimensional column vector whose only non-vanishing entry is 1 at component $j$.\\
Furthermore,
\begin{equation}
  \bar{\tens x}(t) = \mathcal{G}(t,0)\tens x^\mathrm{(i)} - \int_{t_\mathrm{i}}			^t\d t'\,\mathcal{G}(t,t')\tens K(t')\;,
\end{equation}
where
\begin{equation}
  \mathcal{G} = G\otimes\mathcal{I}_N\;,
\end{equation}
is defined with the Green's function $G$ as a $6\times6$ dimensional matrix describing the free propagation of an individual phase-space point.
Just as the phase-space coordinates, the source field $\tens J$ is bundled as
\begin{equation}
  \tens J_q = \vec J_{q_j}\otimes\vec e_j\;,\quad \tens J_p = \vec J_{p_j}				\otimes\vec e_j\;,\quad
  \tens J = \cvector{\vec J_{q_j} \\ \vec J_{p_j}}\otimes\vec e_j\;,
\end{equation}
and for the source field $\tens K$ alike.\\
The interaction between particles is included by applying the interaction part of the action
\begin{equation}
  \hat S_\mathrm{I} = \frac{1}{2}\int\d 1\int\d 2\,\hat\Phi(1)\,\sigma\,(12)\,\hat\Phi(2)
  \label{eq:int_operator}
\end{equation}
to the free action where $\sigma$ is the interaction matrix defined later in (\ref{eq:sigma}) and the operator $\hat\Phi$ bundles the collective field operators for the density $\rho$ and the response field $B$
\begin{equation}
	\hat \Phi := \cvector{\hat \Phi_\rho \\ \hat \Phi_B}\;.
	\label{eq:operator}
\end{equation}
The free generating functional thus contains the collective fields in operator form already, paired with the conjugate source fields $H_\rho$ and $H_B$. The collective field operators themselves are obtained from $Z_0[H,\tens{J},\tens{K}]$ by functional derivatives with respect to the fields $H$,
\begin{equation}
  \hat\Phi_\rho(1) \to \frac{\delta}{\ii\delta H_\rho(1)}\;,\quad
  \hat\Phi_B(1) \to \frac{\delta}{\ii\delta H_B(1)}
\end{equation}
applied to $Z_0[H,\mathrm{J},\mathrm{K}]$.\\
The complete generating functional can thus be written as
\begin{equation}
	\begin{split}
  Z[H,\tens J,\tens K] &= \e^{-\ii\hat S_\mathrm{I}}Z_0[H,\tens J,\tens K]\\ &=\left[1+\left(\e^{-\ii\hat S_\mathrm{I}}-1\right)\right]Z_0[H,\tens J,\tens K]\\
  & \eqqcolon Z_\mathrm{id}+Z_\mathrm{int}
  \end{split}
  \label{eq:full_functional}
\end{equation}
where we have decomposed the full generating functional into the free part for the ideal gas and the interaction part in the last step by effectively adding a zero.
\subsection{Pressure and equation of state}\label{sec: eos}
To later understand how the physical properties like density, pressure and temperature of a gas change due to correlations in the initial conditions we need to relate theses properties to each other by an equation of state. Beginning with the definition familiar from thermodynamics, the pressure is defined as the derivative of the free energy $F$ with respect to the volume while keeping the temperature $T$ and the number of particles $N$ constant
\begin{equation}
	P = -\left(\frac{\partial F}{\partial V}\right)_{N,T}
	\label{eq:pressure}
\end{equation}
where the free energy is given by
\begin{equation}
	F = -\,\mathrm{k_B}T\,\ln{Z_\mathrm{c}}
\end{equation}
with the canonical partition sum $Z_\mathrm{c}$. However, (\ref{eq:pressure}) only holds in thermal equilibrium when the system can be characterized by its macroscopic state variables.
To use this relation for the generating functional derived in Sect. \ref{sec:perturbation} we thus need the system to be in thermal equilibrium. This is ensured since we choose appropriate initial conditions and look at a static situation, i.e.\ at a fixed time $t_0$, without the system having had any time to move out of equilibrium yet. Later, we will assume only weakly correlated initial conditions, making sure that the system will approximately remain in thermal equilibrium.\\
We use (\ref{eq:pressure}) to compute the pressure of a gas and derive an equation of state at an initial time.
We can decompose the full generating functional into the partition sum for the ideal gas $Z_\mathrm{id}$ and for the interaction part $Z_\mathrm{int}$ such that $Z_\mathrm{c}=Z_\mathrm{id}+Z_\mathrm{int}$. For a sufficiently dilute gas with a short ranged interaction potential it is clear that $Z_\mathrm{int} \ll Z_\mathrm{id}$ so that we can approximate the free energy by
\begin{equation}	
	\begin{split}
	&F = -\,\mathrm{k_B}T\,\ln \left(Z_\mathrm{id}+Z_\mathrm{int}\right) = -\,\mathrm{k_B}T\,\ln\left(Z_\mathrm{id}\left(1+\frac{Z_\mathrm{int}}{Z_\mathrm{id}}\right)\right)\\  &\underset{Z_\mathrm{int}\ll Z_\mathrm{id}} 
	\approx\! -\,\mathrm{k_B}T\,\left(\ln Z_\mathrm{id}+\frac{Z_\mathrm{int}}{Z_\mathrm{id}}\right)
	\end{split}
	\label{eq:F}
\end{equation}
We now need to compute $Z_\mathrm{id}$ and $Z_\mathrm{int}$ from the generating functional (\ref{eq:full_functional}).
\subsection{First-order perturbation theory}\label{sec:perturbation}
We treat the functional (\ref{eq:full_functional}) with the Mayer cluster expansion which allows us to expand the interaction potential by the number of particles involved in the interaction. The cluster expansion is based on a clever way to re-write the exponential as
\begin{equation}
	\begin{split}
	&\exp\left(-\beta\sum^N_{i,j=1}v(r_{ij})\right) = \prod^N_{i=1,i< j}\e^{-\beta\,v(r_{ij})}\\
	&= \prod^N_{i=1,i< j} \left[1+f_{ij}\right]= 1+\sum_{i<j}f_{ij}+\sum_{i<j,k<l}f_{ij}f_{kl}+...
	\end{split}
	\label{eq:clusterexp}
\end{equation} 
where we have used the definition of the Mayer function $f_{ij}=\exp\left(-\beta\,v(r_{ij})\right)-1$ with the indices $i,j$ denoting the $i^{\text {th}}$ and $j^\text{th}$ particles.\\
It is clear that in our approach the number of particles taking part in the interaction is encoded in the number of operators $\hat \Phi_{\rho}$ applied to the free generating functional in equation (\ref{eq:genfunc}). We get rid of the factor $\frac{1}{2}$ in front of the sum in the action (\ref{eq:int_operator}) by taking the sum over $i<j$ now. Writing the operators from (\ref{eq:operator}) as a sum over one-particle operators, the exponential we want to expand is given by
\begin{equation}
	\begin{split}
	\e^{-\ii \hat S_\mathrm{I}} &= \exp\left(-\ii\sum^N_{i<j}\int\!\d 1\!\int\!\d 2\,\hat\Phi_i(1)\,\sigma\,(12)\,\hat\Phi_j(2)\right)\\ &= \prod^N_{\substack{i= 1\\i<j}}\exp\left(-\ii \int\d 1\int\d 2\,\hat\Phi_i(1)\,\sigma\,(12)\,\hat\Phi_j(2)\right)
	\end{split}
	\label{eq:interaction_exp}
\end{equation}
with the interaction matrix
\begin{equation}
	\sigma(12) = \matrix{cc}{-\ii v(12)\beta\dirac(t_1-t_i) & v(12) \\ v(12) & 0}\;.
	\label{eq:sigma}
\end{equation}
The $\sigma_{\rho\rho}$ entry is due to the potential energy contribution in the Hamiltonian, 
\begin{equation}
	\mathcal{H}=\sum_i\frac{p^2_i}{2\,m}+\sum_{i<j}v(r_{ij})\;,
	\label{eq:hamiltonian}
\end{equation}
which enters into the initial conditions of a system in thermal equilibrium through the Boltzmann factor.\\ 
The interaction potential is given by 
\begin{equation}
	v(12) = v(\vec q_1 - \vec q_2)\dirac(t_1-t_2)\;.
\end{equation}
We now introduce the Mayer function
\begin{equation}
	\hat f_{ij} := \exp\left(-\ii\int\d 1\int\d 2\,\hat\Phi_i(1)\,\sigma\,(12)\,\hat\Phi_j(2)\right)-1
\end{equation} 
which in our case is in fact an operator with particle labels $i$ and $j$ and can thus expand the exponential (\ref{eq:interaction_exp}) in terms of particle interactions
\begin{equation}
	\begin{split}
	&\prod^N_{\substack{i= 1\\i<j}}\exp\left(-\ii\int\d 1\int\d 2\,\hat\Phi_i(1)\,\sigma\,(12)\,\hat\Phi_j(2)\right)\\ 
	&= 1+\sum_{i<j}\hat f_{ij}+\sum_{i<j,k<l}\hat f_{ij}\hat f_{kl}+...
	\end{split}
	\label{eq:cluster}
\end{equation}
in analogy to (\ref{eq:clusterexp}).\\
To compute the full generating functional (\ref{eq:full_functional}) to first order which means including interactions up to two-particle interactions we have to include only the first and second term on the right hand side of equation (\ref{eq:cluster}). We then obtain the following expression for the generating functional up to two-particle interactions
\begin{equation}
	\begin{split}
	Z^{(1)}= \left[1+\sum_{i<j}\hat f_{ij}\right] Z_0[H,\tens{J},\tens{K}]\Bigr\rvert_{0}\;.
	\end{split}
\end{equation}
Due to the fact that the source fields are responsible for the dynamics of the system and therefore drive the system out of equilibrium we will set them to zero eventually after having taken all required functional derivatives, since our system is supposed to remain in thermal equilibrium.\\
The first term then yields, after setting all source terms to zero,
\begin{equation}
	 Z_\mathrm{id}=Z_0[H,\tens J,\tens K]\Bigr\rvert_{0} = \IC 
	 \label{eq:ideal_sum}
\end{equation}
where the evaluation at zero is a short hand notation for $H=0$, $\tens J =0$ and $\tens K =0$.\\
The second term yields the interaction part of the generating functional,
\begin{equation}
	\begin{split}
	Z^{(1)}_\mathrm{int} &=\sum_{i<j} \hat f_{ij}\, Z_0[H,\tens{J},\tens{K}]\Bigr\rvert_{0}\;.
	\end{split}
	\label{eq:int_sum}
\end{equation}
For convenience, we will neglect the superscript (1) in our further calculations.
\section{Recovery of the classical van der Waals equation of state}\label{sec:vanderWaals}
Since we are interested in the effect correlated initial conditions have on the equation of state of a real gas, therefore we will compute the equation of state of a van der Waals gas in this section and compare it to the equation of state for a real gas with correlated initial conditions that we will compute in Sect. \ref{sec:corrinitialcond}.\\
We now calculate $Z_\mathrm{id}$ and $Z_\mathrm{int}$ for a homogeneous initial density distribution and a Gaussian momentum distribution and recover the equation of state for a classical van der Waals gas.\\
Following the textbook procedure we introduce a potential which describes hard spheres that cannot penetrate each other and feel an attractive potential going as $\vert\vec{q}_{ij}\vert^{-s}$ where $\vert\vec{q}_{ij}\vert$ is the particle distance. Usually $s=6$ is chosen which corresponds to the scaling of the interaction potential between two induced dipoles. The potential is of the form
\begin{align}
v(q) = \begin{dcases*}
        \infty  & for $q<R_0$\\
        -v_0\left(\frac{R_0}{q}\right)^s & for $q>R_0$
        \end{dcases*}
        \label{eq:potential}
\end{align}
where $q:=\vert\vec{q}_{ij}\vert = \vert \vec q_i - \vec q_j\vert$ is the distance between two particles. We see that for particle distances smaller than the particle radius the potential barrier is infinitely high and for distances larger than the particle radius the potential is attractive.\\
For simplicity in further calculations we drop the superscript $(\ii)$ for the initial phase-space coordinates.\\
For the ideal gas term we then obtain 
\begin{equation}
	\begin{split}
	Z_\mathrm{id}&=\int \d\vec q_1\dots \d\vec q_N\int 			\d\vec p_1\dots \d\vec p_N \, \nexp{-\beta \sum_{k}				\frac{\vec p^2_k}{2\,m}} \\ 
	&= V^N\, \left(\frac{2\pi\,m}{\beta}	\right)^{\frac{3N}{2}}
	\end{split}
\end{equation}
where each $\vec q_i$ ranges over the volume $V$ of the gas container.\\
Before evaluating the interaction term we need to integrate over the distance between two particle coordinates $\vec q_{ij} = \vec q_i - \vec q_j$ which leads to the transformation $\d\vec q_j\d\vec q_i = \d\vec q_j\, \d\vec q_{ij}$.
We then have
\begin{widetext}
\begin{equation}
	\begin{split}
Z_\mathrm{int} &= \frac{N(N-1)}{2}V^{N-1}\left(\frac{2\pi\,m}{\beta}	\right)^{\frac{3N}{2}}4\pi\int^{R}_{0}\d \vec q_{ij}\,\hat f_{ij}\, Z'_0[H,\tens{J},\tens{K}]\Bigr\rvert_{0}\\
&\approx - \frac{N^2}{2}V^{N-1}\left(\frac{2\pi\,m}{\beta}	\right)^{\frac{3N}{2}}\!4\pi\left[\int^{R_0}_{0}\!\d \vec q_{ij} -\! \int^{R}_{R_0}\!\d \vec q_{ij}\,\hat f_{ij}\,\right] Z'_0[H,\tens{J},\tens{K}]\Bigr\rvert_{0}
	\end{split}
	\label{eq:Z_int}
\end{equation}
\end{widetext}
where 
\begin{equation}
	Z'_0[H,\tens{J},\tens{K}]= \nexp{\ii \usi t \left\langle{\tens J(t),\tens{\bar x}(t)}\right\rangle}\;.
\end{equation}
We used the approximation that $N(N-1)\approx N^2$ and that $\hat f_{ij}=-1 $ for $v \rightarrow \infty$ going from the first to the second line in (\ref{eq:Z_int}).\\
The first integral in the second line is simply the integration over the particle volume. To evaluate the second integral in the second line we have to expand it to first order in the interaction in order to act with the operators $\hat\Phi$ on the generating functional. After a lengthy calculation presented in appendix (\ref{sec:appendix}) we arrive at the expression
\begin{equation}
	\begin{split}
	\int^{R}_{R_0}\!\d \vec q_{ij}\left(\e^{-\ii \hat S_{ij}}-1\right) Z'_0[H,\tens{J},\tens{K}]\Bigr\rvert_{0}\approx \beta\int \d\vec q_{ij}\, v(\vec{q}_{ij})\;.
	\end{split}
\end{equation}
This is the first-order correction to the ideal gas due to two-particle interactions where the integration ranges over all possible values of the relative position $\vec q_{ij}$, i.e. essentially over the whole volume $V$ of the container.\\
Using the expression derived in (\ref{eq:F}) and the definition of the integral $I\left(\beta\right)$,
\begin{equation}
	I\left(\beta\right) \coloneqq -\,4\pi\left[\int^{R_0}_{0}\!\d \vec q_{ij} -\! \int^{R}_{R_0}\!\d \vec q_{ij}\,\hat f_{ij}\,\right] Z'_0[H,\tens{J},\tens{K}]\Bigr\rvert_{0}\;,
\end{equation}
we can express the free energy as
\begin{equation}
	F = -\,\mathrm{k_B}T\,\left(N\,\ln V + \frac{1}{2}\frac{N^2}{V}I\left(\beta\right)+\frac{3N}{2}\ln\left(\frac{2\pi\,m}{\beta}\right) \right)\;.
\end{equation}
The pressure is then given by
\begin{equation}
	P = \mathrm{k_B}T\,\left(\frac{N}{V}-\frac{1}{2}\frac{N^2}{V^2}I\left(\beta\right)+ \frac{1}{2}\frac{N^2}{V}\frac{\partial}{\partial V}I\left(\beta\right)\right) 
\label{eq:pressurevdw}
\end{equation}
keeping the particle number $N$ and the temperature $T$ constant as described in (\ref{eq:pressure}).
We first consider the last term which usually does not emerge in classical statistical physics with the justification that one considers an infinite volume rather than a finite one and that the separations over which particles can interact are comparably small. This allows to drop the last term completely. We can show that this term is in fact negligible if the shortest length scale on which interactions are possible is much smaller than the volume, even if we consider finite volumes.\\
With $V=R^3$ we thus obtain	
\begin{equation}
\begin{split}
\frac{\partial }{\partial V}I\left(\beta\right)&= - 4\pi\frac{\partial }{\partial V}\left[ \int_{0}^{R_0}\! \d q \, q^2 -\beta\! \int_{R_0}^{R}\! \d q\,q^2\, v_0\left(\frac{R_0}{q}\right)^s \right] \\
 & =  \beta\,v_0 \frac{4\pi}{s-3}\frac{R^3_0}{\alpha^{s-3}}\xrightarrow[\alpha \gg 1]{ }0
\end{split}
\end{equation}
for $s>3$ where we have expressed the radius $R$ in terms of the particle radius $R_0$ as $R=\alpha\,R_0$. Since $R_0$ is of course much smaller than $R$, $\alpha \gg 1$ so that this term can be neglected.
We write (\ref{eq:pressurevdw}) as
 \begin{equation}
	\begin{split}
	P &= \mathrm{k_B}T\, \left[\bar{\rho} - \bar{\rho}^{2}\frac{I\left(\beta\right)}{2}\right] =\mathrm{k_B}T\, \left[\bar{\rho} + \bar{\rho}^{2}B_2(T)\right]
	\end{split}
\end{equation}
with the mean density $\bar {\rho} = \frac{N}{V}$  and  the first virial coefficient $B_2(T)$. 
As a further consistency check we bring this equation into the form of the van der Waals equation of state.\\
For doing so, we integrate the second term in (\ref{eq:pressurevdw}) which yields
\begin{equation}
	\begin{split}
	B_2\left(T\right) &= 2\pi\int_{0}^{R_0} \d q \, q^2 -2\pi\beta \int_{R_0}^{R} \d q\,q^2\, v_0\left(\frac{R_0}{q}\right)^s\\
	&= \frac{2\pi}{3}R^3_0 - \beta\,v_0 \frac{2\pi}{s-3}R^3_0\left(1-\frac{1}{\alpha^{s-3}}\right)\\ &\approx \frac{2\pi}{3}R^3_0 - \beta\,v_0 \frac{2\pi}{s-3}R^3_0\;.
	\end{split}
\end{equation}
Finally we set $s=6$ and define the coefficients $b':=\frac{2\pi}{3}R^3_0$ and $a':=v_0 \frac{2\pi}{3}R^3_0$. The virial coefficient then simplifies to
\begin{equation}
	B_2(T) = b'-\frac{a'}{\mathrm{k_B}T}
\end{equation}
and the pressure is then just
\begin{equation}
	P + \frac{N^2\,a'}{V^2}  = \frac{N \mathrm{k_B}T}{V}\left(1+\frac{N}{V}b'\right)
	 \approx \frac{N \mathrm{k_B}T}{V}\frac{1}{1-\frac{N}{V}b'}
\end{equation}
for $\frac{N}{V}b'\ll 1$. With the amount of substance $n=\frac{N}{N_A}$, the universal gas constant $R=\mathrm{k_B}N_A$ and $a=N^2_A\,a'$, $b=N_A\,b'$, we finally obtain 
\begin{equation}
	\left(P + \frac{n^2\,a}{V^2}\right)\left(V-n\,b\right) = n\,R\,T
\end{equation}
where $a$ measures the attraction between particles and $b$ describes the volume excluded by a mole of particles. This is the well known van der Waals equation of state that can be found in literature.

\section{Correlated initial conditions}\label{sec:corrinitialcond}
\subsection{First virial coefficient for correlated initial conditions}
Adopting the correlated initial conditions derived by \citeauthor{paper_1} \cite{paper_1} we can finally analyse the changes correlations introduce to the equation of state of a non-ideal gas.\\
The probability distribution for the initial position and momentum distribution of particles was derived assuming a statistically homogeneous and isotropic Gaussian random field from which the particles are drawn. The initial velocity field is the gradient of a potential field $\psi$ and thus irrotational. Furthermore, continuity implies that the velocity perturbations have to be coupled to the density such that the velocity potential is the source of the density contrast $\delta$
\begin{equation}
	\delta = - \vec \nabla^2 \psi \quad \text{where}\quad \delta:=\frac{\rho-\bar\rho}{\bar\rho}
	\label{eq:density_contrast}
\end{equation}
with the number density of particles $\rho$ and its mean $\bar\rho$. The power spectrum $P_{\delta}\left(k\right)$ of the density contrast is given by
\begin{equation}
	  \left\langle \hat\delta\left(\vec k\right) \hat\delta\left(\vec k'\right)  \right\rangle = (2\pi)^3 \dirac\left(\vec k + \vec k'\right)P_{\delta}\left(k\right)\;.
\end{equation}
Due to (\ref{eq:density_contrast}) the power spectra for the velocity potential $P_{\psi}\left(k\right)$ and for the density contrast $P_{\delta}\left(k\right)$ must be related by
\begin{equation}
	P_{\psi}\left(k\right) = k^{-4}P_{\delta}\left(k\right)
\end{equation}
so that a single initial density power spectrum is sufficient to describe the complete correlations in phase space.\\
The probability distribution is then given by  
\begin{equation}
  P(\tens q, \tens p) = \frac{1}{\sqrt{(2\pi)^{3N}\det C_{pp}}}\,\mathcal{C}(\tens p\,)
  \exp\left(-\frac{1}{2}\tens p^\top C_{pp}^{-1}\tens p\right)
\end{equation}
where $\tens p = \vec p_j\otimes\vec e_j$ with the correlation operator
\begin{align}
  \mathcal{C}(\tens p) &= (-1)^N\prod_{j=1}^N\left(
    1+\left(C_{xp}\frac{\partial}{\partial\tens p}\right)_j
  \right)\nonumber\\ &+
  (-1)^N\sum_{(j,k)}(C_{xx})_{jk}\prod_{\{l\}'}\left(
    1+\left(C_{xp}\frac{\partial}{\partial\tens p}\right)_l
  \right)\nonumber\\ &+
  (-1)^N\sum_{(j,k)}(C_{xx})_{jk}\sum_{(a,b)'}(C_{xx})_{ab}\prod_{\{l\}''}\left(
    1+\left(C_{xp}\frac{\partial}{\partial\tens p}\right)_l
  \right) + \ldots
\label{eq:correlationmatrix}
\end{align}
where $j\neq k$ as well as $a\neq b$ and $\{\,\}'$ indicates that $l$ runs over all indices except $(j,k)$ and $\{\,\}''$ indicates that $l$ runs over all indices except $(j,k,a,b)$.\\  
The correlation matrices appearing in (\ref{eq:correlationmatrix}) are defined as follows
\begin{align}
  C_{xx} &:= \sigma_2^2\otimes\mathcal{I}_N+\left\langle x_jx_k \right\rangle\otimes E_{jk}\;,\quad
  C_{xp} := \left\langle x_j\vec p_k\right\rangle\otimes E_{jk}\quad\mbox{and}
  \nonumber\\
  C_{pp} &:= \left(\frac{\sigma_1^2}{3}-\frac{\beta}{2\,m}\right)\mathcal{I}_3\otimes\mathcal{I}_N+
  \left\langle\vec p_j\otimes\vec p_k\right\rangle\otimes E_{jk}
\end{align}
with
 \begin{equation}
	\begin{split}
  \left\langle x_jx_k \right\rangle&= \int\frac{\d^3k}{(2\pi)^3}\,P_\delta(k)\,\e^{-\ii\vec k\cdot(\vec q_j-\vec q_k)}\;,\\
  \left\langle x_j\vec p_k\right\rangle &=  \ii\int\frac{\d^3k}{(2\pi)^3}\,k^2\,\vec k\,P_\psi(k)\,\e^{-\ii\vec k\cdot(\vec q_j-\vec q_k)}\;,\\
   \left\langle\vec p_j\otimes\vec p_k\right\rangle&=\int\frac{\d^3k}{(2\pi)^3}\,\vec k\otimes\vec k\,P_\psi(k)\,\e^{-\ii\vec k\cdot(\vec q_j-\vec q_k)}\;.
  	\end{split}
\label{eq: Bmatrix}
\end{equation} 
Note that we include the Boltzmann factor in $C_{pp}$ since we are explicitly considering thermodynamical equilibrium.\\
Assuming that correlations at initial times will be weak, we can approximate the correlation operator to first order in the correlations
\begin{equation}
  \mathcal{C}(\tens d) \approx 1+\sum_{j=1}^N\left(C_{xp}\frac{\partial}{\partial\tens p}\right)_j+
  \sum_{(j,k)}(C_{xx})_{jk}\;.
\end{equation}
Inserting the initial conditions into (\ref{eq:ideal_sum}) we obtain for the first integral 
\begin{equation}
	\begin{split}
	Z_\mathrm{id} =& \int \d\tens q\, \d\tens p\,\frac{1}{\sqrt{(2\pi)^{3N}\det 						C_{pp}}}\,\exp\left(-\frac{1}{2}\tens p^\top 	C_{pp}^{-1}\,\tens p\right)\\ & 			\times \left[1+ \sum_{j}\left(C_{xp}C_{pp}^{-1}\, \tens p\right)_j + 						\sum_{j< k}\xi_{jk}\right]
	\end{split}
\end{equation}
where we abbreviated $\xi_{jk}:=(C_{xx})_{jk}$ with $\xi_{jk}$ being the correlation function.\\
We immediately see that the second integral is a Gaussian integral of the form 
\begin{equation}
	\int^{+\infty}_{-\infty} \d \tens x\, \, \tens x\,\e^{-\frac{1}{2}\tens {x}^{\top} A \tens x} = 0
\label{eq:gauss}
\end{equation}
since $C_{pp}$ and $C_{xp}$ do not depend on $\tens p$ as shown in \citeauthor{paper_1} \cite{paper_1}. Thus the second term is zero which leaves us with the two remaining terms.\\
 We now replace the exponential by its Fourier transform to get rid of the determinant and the inverse of the momentum correlation matrix. With the tensor $\tens t_p := \vec t_{p_j}\otimes\vec e_j$ being the Fourier conjugate to the tensor $\tens p$ we obtain
\begin{equation}
	\begin{split}
	Z_\mathrm{id}=& \int \d\tens q\, \d\tens p\int \frac{\d\tens t_p}{(2\pi)^{3N}}\,\exp\left(-\frac{1}{2}\tens t^{\top}_p C_{pp}\,\tens t_p + \ii \scalar{\tens t_p}{\tens p}\right) \\ &\times \left[1+ \sum_{j < k}\xi_{jk}\right]\;.
	\end{split}
	\label{eq:fourier_trafo}
\end{equation}
Completing the integration over $\tens p$, we see that both terms do not depend on $\tens p$ so that the integration will simply yield a Dirac-delta distribution which, when integrated over $\tens t_p$, sets the exponential function to unity. Since we have chosen the power spectrum such that $P_\delta(0)=0$ we find
\begin{equation}
	\begin{split}
	Z_\mathrm{id} =& \int\!\!\! \d\tens q\int\!\!\!\d\tens t_p\,\exp\left(-\frac{1}{2}\tens t^{\top}_p C_{pp}\,\tens t_p\right)\dirac(\tens t_p)\left(1+ \sum_{j < k}\xi_{jk}\right)\\
	 =& V^{N} + \frac{1}{2} N(N-1)V^{N-2}\,P_\delta(0) = V^{N}\;.
	\end{split}
	\label{eq:id_cor}
\end{equation}
We can now insert the correlated initial conditions into the interaction part of the partition sum (\ref{eq:int_sum}) 
\begin{equation}
	\begin{split}
	Z_\mathrm{int} = &\int \d\tens q\, \d\tens p\,\frac{1}{\sqrt{(2\pi)^{3N}\det C_{pp}}}\,\exp\left(-\frac{1}{2}\tens p^\top C_{pp}^{-1}\,\tens p\right)\\
	& \times\left[1+ \sum_{j}\left(C_{xp}C_{pp}^{-1}\, \tens p\right)_j + 		\sum_{j < k}\xi_{jk}\right]\\
	 &\times\sum_{a<b}\left(\e^{-\ii \hat S_{ab}}-1\right) Z'_0[H,\tens{J},\tens{K}]\Bigr\rvert_{0}\;.
	\end{split}
\end{equation}
After taking the appropriate derivatives the source fields will be set to zero since we consider an equilibrium situation. Then, according to (\ref{eq:gauss}) the second term is zero and we can again apply the Fourier transform to dispose of the inverse matrix $C^{-1}_{pp}$ and its determinant as in (\ref{eq:fourier_trafo}). We then arrive at the expression
\begin{equation}
	\begin{split}
	Z_\mathrm{int} &= \int\!\!\! \d\tens q\,\left(1+\sum_{j < k}\xi_{jk}\right)\sum_{a<b} \hat f_{ab}\, Z'_0[H,\tens{J},\tens{K}]\Bigr\rvert_{0}\\
	&=\frac{N(N-1)}{2}V^{N-1}4\pi\!\!\int^{R}_{0}\!\!\!\d \vec q_{jk}\,\hat f_{ab}\, Z'_0[H,\tens{J},\tens{K}]\Bigr\rvert_{0}\\
	&+\frac{N(N-1)}{2}V^{N-1}4\pi\!\!\int^{R}_{0}\!\!\!\d \vec q_{jk}\,\xi_{jk}\,\hat f_{ab}\,Z'_0[H,\tens{J},\tens{K}]\Bigr\rvert_{0}
\\
&+N(N-1)(N-2)V^{N-2} 4\pi\!\!\int^{R}_{0}\!\!\!\d \vec q_{jk}\,\xi_{jk}\,\hat f_{ab}\, Z'_0[H,\tens{J},\tens{K}]\Bigr\rvert_{0}\;.
	\end{split}
	\label{eq:Zint}
\end{equation}
The second integral in the first line of (\ref{eq:Zint}) contributes three terms due to combinatorics: without loss of generality we set $j=a$ and $k=b$ for the term in the third line and $j=a$ and $k\neq b$ for the term in the last line in (\ref{eq:Zint}). The term for $\{j,k\}\neq\{a,b\}$ vanishes since we have specified $P_{\delta}(0)=0$. The term in the last is several orders of magnitude smaller than all other terms and can therefore be neglected. Using $N(N-1)\approx N^2$ for $N\gg 1$ we can write
\begin{equation}
	Z_\mathrm{int}\approx 	\frac{N^2}{2}V^{N-1}\left[I_1\left(\beta\right)+I_2\left(\beta\right)\right]
	\label{eq:corr_int}
\end{equation}
with the integrals given by 
\begin{equation}
	I_1\left(\beta\right)\approx\frac{N^2}{2}V^{N-1}4\pi\int^{R}_{0}\d \vec q_{jk}\,\hat f_{ab}\, Z'_0[H,\tens{J},\tens{K}]\Bigr\rvert_{0}
\end{equation}
and 
\begin{equation}
	I_2\left(\beta\right)\approx\frac{N^2}{2}V^{N-1}4\pi\int^{R}_{0}\d \vec q_{jk}\,\xi_{jk}\,\hat f_{ab}\, Z'_0[H,\tens{J},\tens{K}]\Bigr\rvert_{0}
\end{equation}
where only $I_2$ contains the spatial correlation function $\xi_{jk}$.\\ 
In the course of this computation we have seen that the momentum correlations do not contribute to the generating functional at all, independent of their order. Returning to (\ref{eq:correlationmatrix}) we see that due to the structure of the terms containing the position-momentum correlations $C_{xp}$ the form of the integrals always reduces to a Gaussian integral of the form (\ref{eq:gauss}) so that the integration yields zero. Thus we see that also the momentum-space correlations do not contribute to $Z^{(1)}$ to any order. This is expected because  the information on the momentum of any particle is only relevant in time. Since we do not consider time evolution, no momentum information can travel between particles.
\subsection{Equation of state for correlated initial conditions}
We now proceed to calculate the pressure for this system by inserting (\ref{eq:id_cor}) and (\ref{eq:corr_int}) into (\ref{eq:F}). The pressure for a correlated van der Waals gas is thus given by
\begin{equation}
	\begin{split}
	P =\mathrm{k_B}T\,\left(\frac{N}{V}-\frac{1}{2}\frac{N^2}{V^2}\left[I_1\left(\beta\right)+I_2\left(\beta\right)\right]\right)\;,
	\end{split}
\label{eq:pressurecorr}
\end{equation}
where the terms containing derivatives of the integrals with respect to the volume  vanish for $R_0\ll V$. Bartelmann et al. \cite{paper_3} computed the density cumulant for correlated initial conditions that we also consider here. Since neither the volume nor the number of particles changes, the mean density has to remain the same independent of correlations or interactions, $\bar{\rho} = \frac{N}{V}$. Equation (\ref{eq:pressurecorr}) as a function of the mean density then reads
\begin{equation}
	P =\mathrm{k_B}T\,\left(\bar\rho-\frac{{\bar\rho}^2}{2}\left[I_1\left(\beta\right)+I_2\left(\beta\right)\right]\right)\;.
\end{equation}
We see that the correlated initial conditions contribute an additional term to the van der Waals gas which depends on the correlation function and therefore on an initial power spectrum.\\ 
We can now evaluate the integrals $I_1\left(\beta\right)$ and $I_2\left(\beta\right)$ for the potential (\ref{eq:potential}) which yields
\begin{equation}
	I_1\left(\beta\right)= -N^2V^{N-1}\left[\frac{4\pi}{3}R^3_0 -\beta\,v_0 \frac{4\pi}{s-3}R^3_0\right]
\end{equation}
and
\begin{equation}
	\begin{split}
	I_2\left(\beta\right)= -N^2V^{N-1}4\pi & \left[\int_{0}^{R_0} \d q \, q^2\, \xi(\vec q)\right.\\
	&\left. -\beta\,v_0\! \int_{R_0}^{R} \d q\,q^2\,\xi(\vec q)\left(\frac{R_0}{q}\right)^6\right]
	\end{split}
\end{equation}
where we have expanded to first order in the interaction potential.\\
We see that the contribution due to spatial correlations has a term of the same order in $N$ and $V$ as the contribution due to the interaction itself.\\
\subsection{Influences of correlations for a simple model}
To determine how the initial correlations affect the pressure of a gas we compute the second integral in (\ref{eq:pressurecorr}). As a toy model, we use a simple power law for the correlation function 
\begin{align}
\xi(q) = \begin{dcases*}
        0 & for $q<R_0$\\
       A\,\left(\frac{q}{q_0}\right)^{-\gamma} & for $q>R_0$\;.
        \end{dcases*}
        \label{eq:corfunction}
\end{align}
The correlation function $\xi(q)$ is zero for $q<R_0$ because otherwise we would include the repulsive potential of the hard spheres twice. We chose $q_0$ to be the mean particle separation. The amplitude of the correlations is set in such a way as to increase the probability $\d P \propto [1+\xi(q_0)] $ for finding a particle at a distance $q_0$ from another by a factor to be set.\\
\begin{figure}
	\includegraphics[scale=0.45]{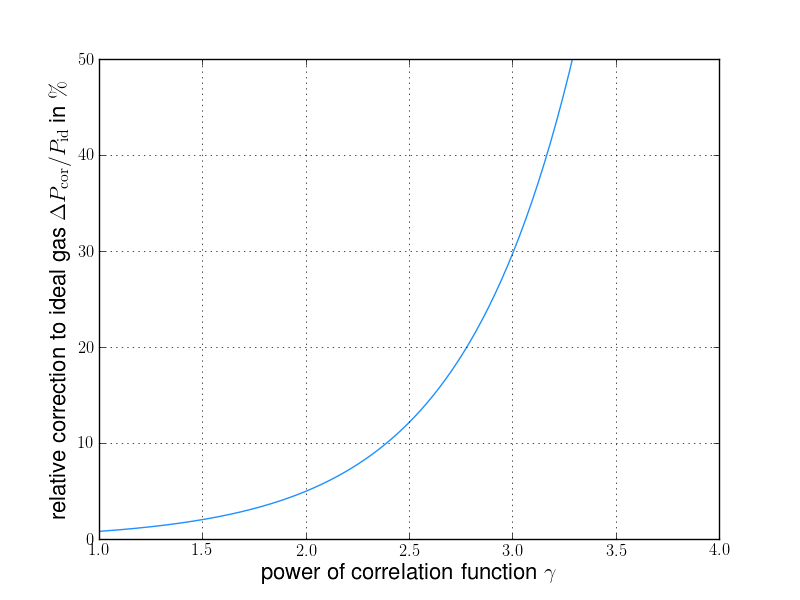}
	\caption{The dependence of the relative pressure correction to the ideal gas due to correlations on the power of the correlation function $\gamma$ is shown in percent. The pressure contributions to the ideal gas pressure from the correlations rapidly increase with the power of the correlation function.}
	\label{fig:rel_pressure}
\end{figure}
The correction due to correlated initial conditions to the pressure of an ideal gas is then given by
\begin{equation}
	\Delta P_{\mathrm{cor}} = -2\pi\,\frac{N^2}{V^2}A\,q^{\gamma}_0\frac{v_0}{3+\gamma}R^{3-\gamma}_0\;.
\end{equation}
For a diatomic gas (e.g. $\mathrm{N_2}$ with $v_0 \approx 10^{-21}$ taken from \cite{Lennard-Jones}) at a temperature of $T=273\mathrm{K}$ with $N=10^{27}$ particles at a volume of $V=22.4\, \mathrm{m^3}$ we can compute the correction due to the van der Waals interaction to the ideal gas which increases the gas pressure by about $0.5\%$. The corrections are quite small. Most importantly, the repulsive term dominates so that the pressure increases. If we consider a sufficiently deep potential well we can also obtain negative pressure corrections to the ideal gas when the attractive term starts to dominate. If we  now switch on correlations with the correlation power $\gamma=2$ and a correlation amplitude of $A=1$, thus doubling the probability to find a particle at the mean particle separation $q_0$,  the pressure corrections become negative. Reaching $5\%$, the corrections due to correlations are an order of magnitude larger than the corrections resulting from van der Waals interactions. The the distribution of the gas particles is clumpy and therefore the overall pressure decreases. From (\ref{eq:corfunction}) we can see that the correction to the pressure of the ideal gas linearly increases with the amplitude $A$ of the correlation function. In Fig. \ref{fig:rel_pressure} we show the dependence of the relative pressure corrections due to correlations to the pressure of an ideal gas. We see that with higher power of the correlation function, the relative pressure correction rapidly increases. However, if the correlations become too strong our assumption about the system being in thermal equilibrium will no longer hold.

\section{Conclusions}\label{sec:conclusion}
We have used the formalism first proposed by \citeauthor{2010PhRvE..81f1102M} and \citeauthor{2013JSP...152..159D} \cite{2013JSP...152..159D, 2012JSP...149..643D, 2011PhRvE..83d1125M, 2010PhRvE..81f1102M} and adopted the correlated initial conditions by \citeauthor{paper_1} \cite{paper_1} to investigate the impact of initial correlations on van der Waals gases by calculating the equation of state. We have first conducted a consistency check for the approach and derived the van der Waals equation of state and then imposed correlations to qualitatively and quantitatively study their influence on the gas. Our main results are:
\begin{itemize}
	\item Since we are considering a system in thermal equilibrium and can therefore set all source terms to zero, the only correlations left to consider are spatial correlations. Correlations containing the momentum of particles do not appear in our static situation, since the information on the momentum of a particle is only relevant in time.
	\item By considering Gaussian momentum and spatially homogeneous initial conditions we can easily reproduce the van der Waals equation of state known from classical thermodynamics describing the correction to an ideal gas due to short-ranged particle interactions.
	\item The imprinted correlations cause a pressure decrease in the gas which is an order of magnitude larger than the correction due to short-ranged particle interactions for the correlation function we chose. This significant correction must thus be included in studies of non-ideal gases with initially correlated particles.
\end{itemize}
We have now obtained a tool to study systems which do not only contain an interaction potential, but also correlated initial conditions. These systems can be studied in laboratories already and show very interesting behaviour considering phase transitions. Therefore one of the goals for future studies is to describe such phase transitions analytically with this new approach.

\section{Acknowledgments}
We would like to thank Daniel Berg and Björn Schäfer for helpful and inspiring discussions.

\section{Appendix}
\subsection{Computation of the interaction term to first order}\label{sec:appendix}
In our later calculation we will need to perform the integral over the initial conditions. For this purpose, we will need to expand the second term in the interaction potential which leads to the expression
\begin{equation}
	\begin{split}
  	&\int^{R}_{R_0}\!\d \vec q_{ij}\left(\e^{-\ii\int\d 1\int\d 2\,\hat\Phi_i(1)\,\sigma\,(12)\,\hat\Phi_j(2)}-1\right) Z'_0[H,\tens{J},\tens{K}]\Bigr\rvert_{0}\\
  	\approx & -\,\ii\int^{R}_{R_0}\!\d \vec q_{ij}\left(\int\!\!\d 1\!\!\int\!\!\d 2\,\der{}{H_\rho(1)}\,\sigma_{\rho B}\,(12)\,\der{}{H_B(2)}\right.\\
  	&\left.- \int\!\!\d 1\!\!\int\!\!\d 2\,\der{}{H_\rho(1)}\,\sigma_{\rho\rho}\,(12)\,\der{}{H_\rho(2)}\right) Z'_0[H,\tens{J},\tens{K}]\Bigr\rvert_{0}\\
  	&\eqqcolon  T_1+T_2\;.
 	\end{split}
\end{equation}
For the evaluation of this term we transform to Fourier space for simplicity. 
The Fourier transform of the interaction potential to k-space yields
\begin{equation}
	\begin{split}
	&\frac{1}{\left(2\pi\right)^6}\int \d \vec x_1 \d \vec x_2 \int \d \vec k_1\d \vec k_2 \, v(\vec x_1 - \vec x_2) \e^{-\ii \vec k_1\cdot \vec x_1}\e^{-\ii \vec k_2\cdot \vec x_2}\\ &= \frac{1}{\left(2\pi\right)^3}\int \d \vec k_1\d \vec k_2 \, v\left(\vec k_1\right)\dirac\left(\vec k_1+\vec k_2\right)\;.
	\end{split}
\end{equation}
The first term excluding the integration over the initial conditions then reads 
\begin{widetext}
\begin{equation}
	\begin{split}
	T_1 =&\, -\ii\! \int^{R}_{R_0}\!\d \vec q_{ij}\int \d t_1 \d t_2\,\dirac(t_1-t_2)\int \frac{\d\vec 			k_1 \d\vec k_2}{\left(2\pi\right)^3}\, \der{}{H_{B_{i}}(t_2, -\vec k_2)}\,v(\vec k_1)\dirac(\vec k_1+\vec k_2)\\ 
	&\times\der{}{H_{{\rho}_{j}}(t_1, -\vec k_1)}\,\nexp{\ii\sum_{a=\rho,B}				\sum^N_{l=1}\int \frac{\d t'_1 \d\vec k'_1}{\left(2\pi\right)^3} H_{a_{l}}(t'_1,\vec k'_1)\,						\hat{\Phi}_{a_{l}}(t'_1, -\vec k'_1)}Z'_0[\tens{J},\tens{K}]\Bigr\rvert_{0}
	\end{split}
\end{equation}
\end{widetext}
where $Z'_0[\tens J,\tens K]$ denotes the generating functional excluding the integration over initial conditions. The operators $\hat\Phi_{a_j}$ are given by
\begin{equation}
	\begin{split}
	\hat\Phi_{\rho_j}(1) &= \exp\left(-\vec k_1^{\,\top}\cdot\frac{\delta}					{\delta\vec J_{q_j}(1)}\right) \hspace{0.5cm}\text{and}\\ 							  \hat\Phi_{B_j}(1) &= \left(\vec k_1^{\,\top}\cdot						\frac{\delta}{\delta\vec K_{p_j}(1)}\right)\hat\Phi_{\rho_j}(1) =: \hat 					b_j(1)\hat\Phi_{\rho_j}(1)\;.
	\end{split}
\end{equation}
Applying the functional derivatives and already setting $H$ to zero we obtain
\begin{equation}
	\begin{split}
	T_1 = - \ii\!\int^{R}_{R_0}\!\!\!\!\d \vec q_{ij}\!\!\int\!\d t_1\!\!\!\int\!\! \frac{\d\vec k_1}{\left(2\pi\right)^3}\,& \hat\Phi_{B_{i}}(t_1, \vec k_1)\,v(\vec k_1)\\ &\times\hat\Phi_{\rho_{j}}(t_1, -	\vec k_1)\,Z'_0[\mathrm{\tens J},\mathrm{\tens K}]\Bigr\rvert_{0}
	\end{split}
\end{equation}
where we have already integrated out the Dirac-delta distributions. The action of the density operator on the generating functional will result in a shift of the tensor $\tens{J}$ by a tensor $\tens{L}_j$ which is defined by
\begin{equation}
	\begin{split}
 	\tens L_j(1) &:= -\vec k_1\cdot\frac{\delta\tens J(t)}{\delta\vec J_{q_j}(1)} 			= -\vec k_1\cdot\cvector{\delta_\mathrm{D}(t,t_1)\mathcal{I}_3 \\ 0_3}			\otimes\vec e_j\\ & = -\delta_\mathrm{D}\left(t-t_1\right)\cvector{\vec k_1\\ 			0}\otimes\vec e_j
 	\end{split}
\end{equation}
as discussed in  \citeauthor{paper_1} \cite{paper_1}. We further introduce the shorthand notation
$\tilde k_1 := \cvector{\vec k_1\\ 0}$, so that after applying the density operators and setting $\tens J$ to zero we arrive at
\begin{equation}
	\begin{split}
	T_1 =& \,-\ii\int^{R}_{R_0}\!\d \vec q_{ij}\int\! \d t_1 \int \frac{\d\vec k_1}{\left(2\pi\right)^3}\, v(\vec k_1)\hat b_j(t_1, \vec k_1)\\&\times\nexp{\ii\usi t \left[\tens L_i(t_1, \vec k_1)\bar{\tens x}_{q_i}-\tens L_j(t_1, \vec k_1)\bar{\tens x}_{q_j}\right]}\Bigr\rvert_{\tens{K}=0}\;.
	\end{split}
\end{equation}
Applying the derivative with respect to the source field $\vec K_{p_i}$ and setting $\tens K$ to zero leads to
\begin{widetext}
\begin{equation}
	\begin{split}
	T_1 =& \int^{R}_{R_0}\!\d \vec q_{ij}\int \d t_1\,\int\frac{\d\vec k_1}{\left(2\pi\right)^3}\, v(\vec 	k_1) \,\nexp{\ii\usi t \left[\sum_{b=q,p}\,\left(\tens L_i(t_1, \vec k_1)g_{qb}(t,t_0)\bar{\tens x}^{(\ii)}_{b_i}  
	-\tens L_j(t_1, \vec k_1)g_{qb}(t,t_0)\bar{\tens x}^{(\ii)}			_{b_j}\right)\right]}\\ 
	&\times \vec k_1\usi t \left(\tens L_i(t_1, \vec k_1)\usi 	t' g_{qp}(t,t')\dirac(t'-t_1)\delta_{ij} - \tens L_j(t_1, \vec k_1) \usi t' g_{qp}(t,t')\dirac(t'-t_1)\right)=0 \\
	\end{split}
\end{equation}
\end{widetext}
where we have used that $g_{qp}(t,t)=0$ in the last step implying that $T_1$ will vanish.\\
In analogy to the previous calculation we obtain an expression for the second term after setting all source fields to zero
and using that $g_{qq}(t,t)=1$ and $g_{qp}(t,t)=0$. The only surviving term is then 
\begin{equation}
	\begin{split}
	T_2 = \beta\int^{R}_{R_0}\!\d \vec q_{ij} v(\vec{q}^{(\ii)}_i - \vec{q}^{(\ii)}_j)\;.
	\end{split}
\end{equation}
We thus arrive at the expression for the part of the partition function containing the attractive potential which we have used in the computation of the first virial coefficient for two different sets of initial conditions.

\bibliography{my_bib}
\end{document}